# Biomimetic emulsions reveal the effect of homeostatic pressure on cell-cell adhesion

Lea-Laetitia Pontani[1], Ivane Jorjadze[1], Virgile Viasnoff[2], Jasna Brujic[1]

[1]New York University, Department of Physics and Center for Soft Matter Research,
4 Washington Place, New York, NY, 10003, USA
[2]Mechanobiology Institute, Singapore

To whom correspondence should be addressed; E-mail: jb2929@nyu.edu.

## Abstract
Cell-cell contacts in tissues are continuously subject to mechanical forces due to homeostatic pressure and active cytoskeleton dynamics. While much is known about the molecular pathways of adhesion, the role of mechanics is less well understood. To isolate the role of pressure we present a dense packing of functionalized emulsion droplets in which surface interactions are tuned to mimic those of real cells. By visualizing the microstructure in 3D we find that a threshold compression force is necessary to overcome electrostatic repulsion and surface elasticity and establish protein-mediated adhesion. Varying the droplet interaction potential maps out a phase diagram for adhesion as a function of force and salt concentration. Remarkably, fitting the data with our theoretical model predicts binder concentrations in the adhesion areas that are similar to those found in real cells. Moreover, we quantify the adhesion size dependence on the applied force and thus reveal adhesion strengthening with increasing homeostatic pressure even in the absence of active cellular processes. This biomimetic approach reveals the physical origin of pressure-sensitive adhesion and its strength across cell-cell junctions.

## Introduction

Cell-cell adhesion is important in biology because it underlies the structure of tissues and their dynamic reorganization during processes as important as morphogenesis [1, 2], cell locomotion [3, 4] and signaling [5, 6]. In addition to the high level of complexity in the identified biochemical pathways, it has recently become clear that mechanical effects also play an important role. For example, pushing cells together or increasing their contractile forces by changing the substrate stiffness reinforces the strength of contacts [7, 8, 9]. Furthermore, since homeostatic pressure arising from the balance of cell division and cell death is important in achieving the mechanical integrity of tissues [10] it should also affect cell-cell adhesion. Despite these important observations, the physical origin of force-sensitive adhesion remains an open question. In fact, theoretical models are derived from the behavior of simplified model membranes that lack mechanical resilience [11]. Although these models successfully describe the kinetics and energetics of adhesion in the absence of rigidity [12, 13], they cannot address the effects of force. In a cell, rigidity arises from the cytoskeleton scaffold and the mechanical coupling with neighboring cells in the surrounding tissue. As a result, individual cells are viscoelastic with a bulk modulus of ~1kPa [14]. Moreover, the interplay between cortical tension and adhesive interactions with neighbors gives rise to a surface tension in cellular aggregates [15, 16].

Here we present a biomimetic emulsion system, in which the elasticity is introduced through an interfacial tension of _10mN/m to match the one measured in cell aggregates and embryonic tissues [17, 18]. Furthermore, to mimic the dense packing of cells in tissue we compress the 3D assembly of droplets at _10kPa, in good agreement with the measured homeostatic pressure in tissues [19, 20]. In addition to mechanical similarities, the chemical composition of the emulsion system reproduces the attractive and repulsive interactions that govern adhesion between cells. By experimentally tuning the interaction potential and the elasticity of the emulsion we show the conditions under which a pushing force is necessary to create adhesion, which lends credence to the hypothesis that actin-mediated forces are a prerequisite for cell-cell adhesion. The dependence of this threshold force on the interparticle interaction is captured by a free energy model, which results in a phase diagram in agreement with our experiments.

## Results and Discussion

### Biomimetic emulsions

In the model system illustrated in Figure 1A, the emulsion is stabilized by a mixture of surface active agents that serve different biomimetic purposes: EPC phospholipids are the major interfacial component that replace the outer leaflet of cell membranes; SDS ionic surfactant introduces electrostatic repulsion between the droplet surfaces, analogous to the charge repulsion between cell surface macromolecules; PEG-biotinylated lipids act as a polymer brush to induce steric repulsion [21] and also provide biotin ligands for the binding with streptavidin to mimic adhesion mediated by homophilic cadherins. A biotin-streptavidin complex on one droplet surface diffuses until it binds to another biotin on a neighboring surface through a second binding site, which is the molecular basis for adhesion in this model system [22, 23]. The energy associated with forming such a ligand-receptor bond is on the same order as that of cellular adhesive junctions [24, 25, 26]. In order to visualize the droplet assembly in 3D shown in Figure 1B we match the refractive indices of the aqueous and oil phases and use fluorescent streptavidin to label the biotinylated lipids on the droplet surface. The homogeneous fluorescence of the lipids at the interface [27] provides a quantitative measure of the concentration of PEG-biotinylated lipids. In this case, the equilibrium separation between the droplet surfaces exceeds the distance over which the proteins can interact and the fluorescence remains homogeneous. We therefore compress the droplets by centrifugation at rates ranging from 1 to 1000 times the acceleration due to gravity to deform them such that their surfaces are closer together. A given centrifugation rate gives rise to a broad distribution of deformation areas corresponding to a distribution of interdroplet forces within the 3D packing [28]. To extend the force range even further and increase the statistical accuracy of our results we centrifuge each emulsion at different rates and pool together data from different 3D stacks. When the interdroplet force leads to a surface-to-surface distance that allows for protein interactions, we observe a redistribution of the proteins into adhesive patches that give rise to a much higher fluorescence intensity at the center of the droplet deformation sites shown in Figure 1C.

**Model for adhesion between droplets**

To quantitatively assess adhesion in the emulsion system, we develop a local model that takes into account the various energy terms that play a role in the interaction between two droplets in contact, as illustrated in the schematic in Figure 2A. The binding energy $E_b$ of the lock and key proteins and the work $W_l$ done by the external pressure both favor adhesion. By contrast, the electrostatic repulsion $E_e$ and the surface energy $E_d$ oppose adhesive patch formation. Van der Waals interactions, which are negligible in cell-cell contacts, are ignored since the system is refractive index matched. For a given set of experimental conditions, the minimization of the total free energy $E = E_d + E_e + E_b + W_l$ with respect to the distance h between the droplet surfaces and the deformation angle $\theta$ sets the equilibrium adhesion patch size, as shown in Figure 2A.

For clarity, we describe a simplified model that neglects emulsion polydispersity and volume conservation upon deformation, which are relaxed in the Supplementary Materials. The work done by the external pressure on the droplets is given by $W_l = Fd$, where F is the interdroplet force and $d = R_0(\frac{h}{R_0} + 2 - \theta + \frac{5}{24}\theta^4)$ is the distance between the droplet centers and $R_0$ is their undeformed radius. This compression of the droplets does work against the energy of deformation given by the Princen model as $E_d = \frac{1}{2\sigma \pi R_0^2 \theta^4}$, where $\sigma$ is the surface tension. Moreover, the work done in bringing the surfaces closer together serves to overcome electrostatic repulsion, modeled as $E_e = 2\pi\varepsilon\psi_0^2 R_0 \exp(-\kappa h)$ where $\varepsilon$ is the dielectric constant, $\psi_0$ is the electrical potential at the droplet surface, and $\kappa$ is the inverse of the Debye length. If the resulting interparticle distance h is smaller than a critical length $h_c$, the interaction of the surface proteins leads to an adhesive state with an additional binding energy term. The distance $h_c$ of about 18nm is set by the size of the biotinylated lipids and the streptavidin between them. The binding energy is given by $E_b = e_b R_0^2 (\theta^2 - \frac{1}{3}\theta^4) H(h_c - h)$, where $e_b = c_b \varepsilon_b$ is the binding energy per unit area, $c_b$ is the binder concentration in the adhesion patch, $\varepsilon_b$ is the binding energy of an individual binder [29] and H(t) is the Heaviside function that determines whether binding is allowed. The resulting energy landscape reveals two local minima, $E_1 = E(\theta_1; h_1)$ and $E_2 = E(\theta_2; h_2)$, corresponding to the deformed yet non-adhesive and adhesive states of the contacting droplets. These energy states are separated by an energy barrier and a discontinuity at $h = h_c$ that comes from the additional binding energy $E_b$ for $h < h_c$, as shown in figure 2B. Adhesion can only occur if the global minimum is found at $h < h_c$ or if the energy difference between the two minima is within thermal energy. Decreasing the Debye length or compressing the droplets with force F modifies the energy landscape in favor of the adhesive state, as shown in Figure 2C. These parameters are varied in the experiment to probe the validity of the model.

## Quantitative analysis of force-dependent adhesion

In order to compare the experimental findings with the model we first extract ~1000 adhesion patches from 3D reconstructions of confocal images. As shown in Figure 3A, the adhesion patches are identified by a thresholding algorithm because they fluorescence brighter than the surface of the droplets or the aqueous background (see Supplementary Materials). The homogeneous spatial distribution of the adhesion discs within the volume of the packing is shown in Figure 3B in a typical experiment. For every droplet pair in contact we therefore measure the adhesion patch radius $r_p$. Second, the images reveal the radius of deformation $r_d$ between contacting droplets from the geometric overlap between the identified spheres of radius R shown in Figure 3C [30]. Whereas the adhesion patch spans the full area of deformation in the theoretical model, experimentally we observe $r_p < r_d$. We find that $r_p = \alpha r_d$ with the slope $\alpha$ giving the coverage of the adhesion (Figure 3D). We use the measured $r_p$, $r_d$, R, and the model parameters introduced above (see Supplementary Materials for values) to obtain the work done in compressing each droplet pair $W_l = E_e + E_d - E_b$ and the corresponding interdroplet force $F_l$.

This analysis allows us to test how electrostatic repulsion, surface tension, and the screening of charges influence the force-dependent adhesion in terms of the timescale, size and number density of the protein links. By varying the SDS concentration from 1 to 5mM in the emulsions, we simultaneously increase the charge repulsion and decrease the surface tension of the droplets. In the 5mM case, the charge repulsion prevents adhesion under gravity (Figure 4Ai) and requires an applied pressure by centrifugation as well as a long waiting time before the patches form (Figure 4Aii). The fact that patches persist after relaxing the applied pressure to 0.2kPa, corresponding to gravitational compression, confirms that they arise from protein links across contacting surfaces (Figure 4Aiii). This irreversibility indicates a kinetic barrier to removing the adhesive patches. The mean patch radius grows towards steady state size to form adhesions on a characteristic timescale of hours, as shown in Figure 4B, where the patch growth dynamics is displayed for two different global pressures. Interestingly, these timescales on the order of hours are significantly slower than minutes encountered in individual cellular adhesions [31] or seconds in functionalized model membranes [32]. However, centrifugation-based bulk measurements of the kinetics of cell-cell adhesion reach a plateau after 90 minutes [33], similar to the ~120 minutes measured under the low emulsion compression. Decreasing the electrostatic repulsion by lowering the SDS concentration to 1mM or by screening charges with salt leads to patches growing on much faster timescales (below 20 minutes), independent of the centrifugation rate (Figure 4Cii-vi).

Image analysis of the local microstructure reveals the dependence of each patch size on the corresponding interdroplet force. To probe a wide range of forces we centrifuge each emulsion at different rates and image multiple stacks to collect a large statistical pool of data. We find that higher compression visibly increases the adhesion patch sizes under all conditions (Figure 4Ciii,vi). To quantify this effect we bin the local interdroplet forces and plot the corresponding average patch size as a function of the average force for all conditions, as shown in Figure 4D. In all cases, the increase of patch size with load force follows the model prediction of a square root law at high forces, but there is a pronounced deviation towards larger patches at low forces due to the onset of protein binding. This result suggests that mechanical compression is sufficient to induce cell-cell adhesion strengthening, in addition to the active forces exerted by actin polymerization [9]. While the force-dependence is similar between the data sets, they differ in the prefactor. This prefactor corresponds to the adhesion coverage α of the area of deformation identified in Figure 3D, which is larger for the 5mM SDS emulsion with salt. A possible explanation is that a line tension develops as the protein complexes displace the other surface molecules and increase the local surface tension, similar to domain formation. The coverage efficiency then depends on the surface properties of the emulsion and the resulting line tension. In cell-cell adhesion, such a line tension could account for the initial cadherin accumulation into small puncta that spread across the interface over time [34].

Although the increase of patch sizes with force follows the model prediction independent of the emulsion conditions, the fraction of droplets contacts that are covered with adhesion patches $N_p=N_c$ reveals interesting distinctions, as shown in Figure 5A. In the absence of screening by salt no patches are observed in the 1mM and 5mM SDS emulsion under gravitational compression with forces of ~15pN (corresponding to deformations below the resolution limit of the microscope). On the other hand, applying the smallest measurable force of ~2nN leads to 20% and 35% of droplet contacts with adhesions, respectively. This result is consistent with the force tilting the energy landscape in the model to favor the adhesive minimum. However, the low probability of adhesion remains constant over the entire force range up to 50nN, which indicates a kinetic barrier that is insensitive to force. Instead, this barrier can be overcome by screening the electrostatic repulsion with 10mM salt, which allows some adhesions (5%) to form even under gravity. Upon compression of the screened emulsions the probability of adhesion reaches almost 1, also evidenced by the large number of patches in Figs. 4Cv-vi. The effects of these experimental scenarios on the model are shown in Figure 2C for the 5mM SDS emulsion. They highlight the importance of homeostatic pressure in achieving the mechanical integrity of tissues [10]. An alternative to using force to populate droplet contacts with adhesions is to screen the charges by increasing the salt concentration, as shown in Figure 5B and the image in Figure 4Civ. This trend is in agreement with the model, in which the corresponding decrease in the Debye length changes the energy landscape, favoring the adhesive state and decreases the barrier to it. Since the model assumes a constant compression force of 15pN between droplets, the transition appears sharper than in the emulsion where the patch fraction is derived from a distribution of forces in a given droplet packing under gravity. Under physiological conditions of 100mM salt the model predicts the spontaneous nucleation of adhesions in both emulsions. Under gravity alone, the model predicts adhesions on the scale of 200nm in radius from the estimated concentration of cadherins on the cell surface [16, 35]. While such small adhesions are sufficient to trigger a biochemical response in the cell, they cannot maintain the mechanical integrity of tissues. As shown above, nanoNewton forces are necessary to grow adhesions that span the entire cell-cell interface.

**Phase diagram for adhesion**
Finally, we construct phase diagrams for adhesion from the probabilities of forming a patch as a function of the applied force and the concentration of NaCl, as shown in Figure 5C for the 5mM SDS emulsion. The model prediction of the phase diagram, fixed by literature values for the surface tension of our emulsions, the binding energy per streptavidin-biotin bond and the measured value for the electrical potential, yields a binder concentration of 47 molecules/µm$^2$ in the 5mM SDS case and 60 molecules/µm$^2$ in the 1mM SDS case to fit the phase boundaries identified by the data. Remarkably, this range of binder densities is similar to that of cadherins (80-800 molecules/µm$^2$) on the cell surface. The fact that all the parameters that describe the biomimetic system are to within a factor of two in agreement with the values measured in cells under physiological conditions lends strong support to this synthetic approach in biology. The predicted boundary between the adhesive and non-adhesive regions in phase space explains why cell aggregates either spread like a viscoelastic medium or disperse like an assembly of particles depending on the adhesion properties of the cell-cell interactions[36]. While it is known that the concentration of adhesive molecules on the cell surface tune the strength of adhesion, our model system shows how this concentration depends on the balance between factors such as the osmolyte concentration, membrane surface tension and cytoskeletal pushing forces as well. More specifically, the phase diagram demonstrates that the global screening of charges present in the cellular environment facilitates the formation of adhesions, but external pressure is necessary to strengthen the cellular interfaces for tissue integrity. This result highlights the possible role of adhesion in tumor progression, since homeostatic pressure affects its growth and metastasis. The versatility of our simplified system will enable the quantitative study of specific constituents in the mechano-sensitive regulation of cellular adhesion.

## Materials and Methods

The protocol for the emulsion preparation is inspired by experiments described in [27]. Here, the oil droplets contain egg L-α-phosphatidylcholine (EPC) lipids and the DSPEPEG(2000) biotinylated lipids from Avanti Polar Lipids (Alabaster, AL) at a molar ratio of 92:8, respectively, and a total mass of 19mg. The solvent containing the lipids is evaporated under nitrogen before 10mL of silicone oil is added to the dried lipids. This mixture is then sonicated during 30 minutes at room temperature and heated at 50₋C during 3 hours. After cooling to room temperature the lipid containing oil (10mL) is first coarsely emulsified in 22mL of buffer (5mM SDS, w$_t$ = 18% dextran). This crude emulsion is then injected into a narrow gap couette mixer, with a gap size of 100 µm, and sheared at 22rpm. The resulting emulsion is washed twice in an aqueous solution of 1 or 5 mM SDS before a last wash in the index matching buffer containing 50:50 glycerol:water. This emulsion is stable over several weeks at 4ᵒC. The emulsion is dyed on the surface with Texas Red conjugated streptavidin (Invitrogen), 500 µL of 1 or 5 mM SDS emulsion is mixed with 1mg/mL streptavidin (25₋L) and 1.5mL of buffers containing 2mM Tris pH=7, 1 or 5mM SDS, 0 to 30mM NaCl. This solution is incubated 1h at room temperature to allow the streptavidin to bind to the biotinylated lipids on the droplets. The sample can be observed after creaming under gravity as shown in Figure 1B or centrifuged a 20 ᵒC at accelerations ranging from 50 to 1400g during 20 minutes. The top layer of the compressed emulsion is then transferred into another observation cell to isolate it from the continuous phase and therefore avoid relaxation. The sample is imaged using a fast scanning confocal microscope (Leica TCS SP5 II).

**Acknowledgments**

J.B. acknowledges support from US NSF Career Award 0955621. L.-L. P. was supported in part by New York University Materials Research Science and Engineering Center Award DMR-0820341. We thank Eric Vanden Eijnden and Izabela Raczkowska for fruitful discussions and David Pine and Paul Chaikin for a careful reading of the manuscript.

14

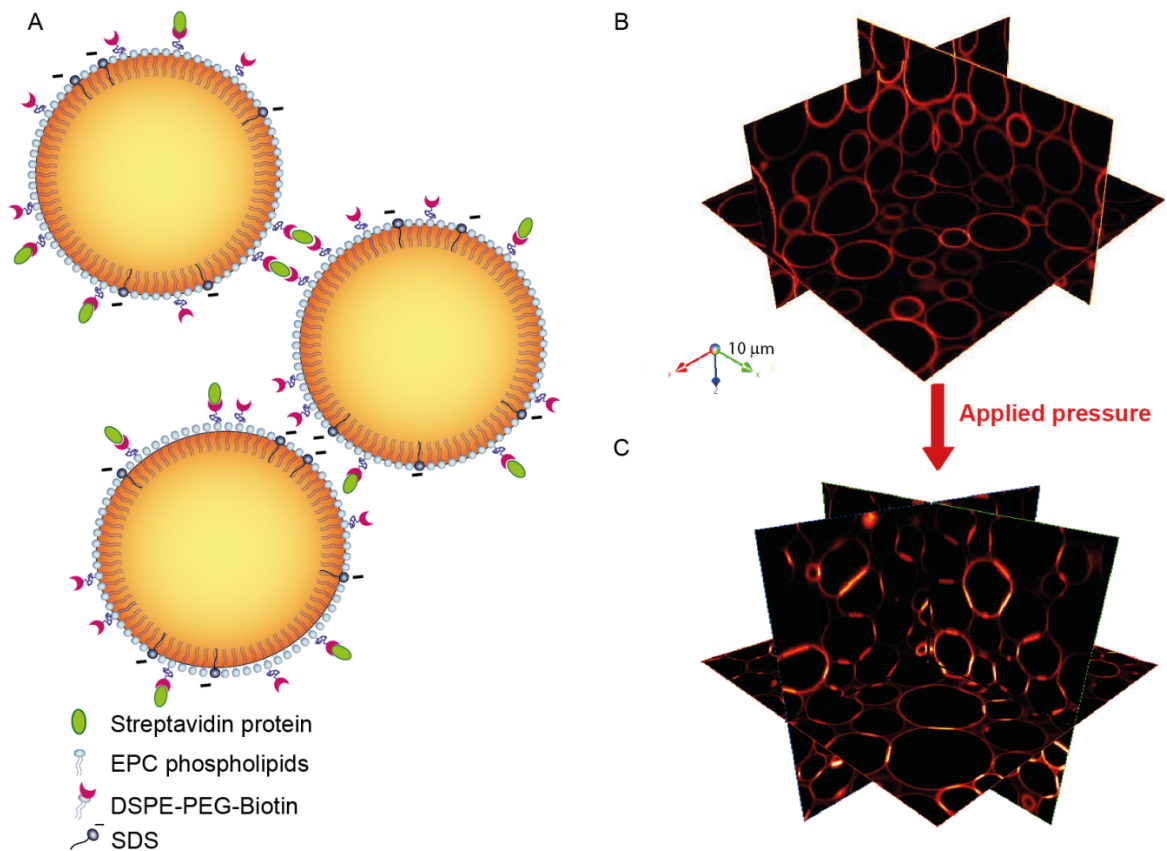

**Figure 1** (A) Schematic representation of functionalized emulsions. The oil/water interface is stabilized by a mixture of phospholipids and negatively charged SDS. Some of the lipids hold a PEG-biotin group that allows binding through biotin-streptavidin interactions, as shown on the upper droplets. (B) Three-dimensional representation of confocal images shows Texas Red-streptavidin fluorescence on the surface of the droplets. Packing under gravity is not sufficient to create adhesion between the droplets (top image), whereas an applied pressure triggers the formation of adhesions between the droplets, shown as the areas of brighter fluorescence in the bottom image.

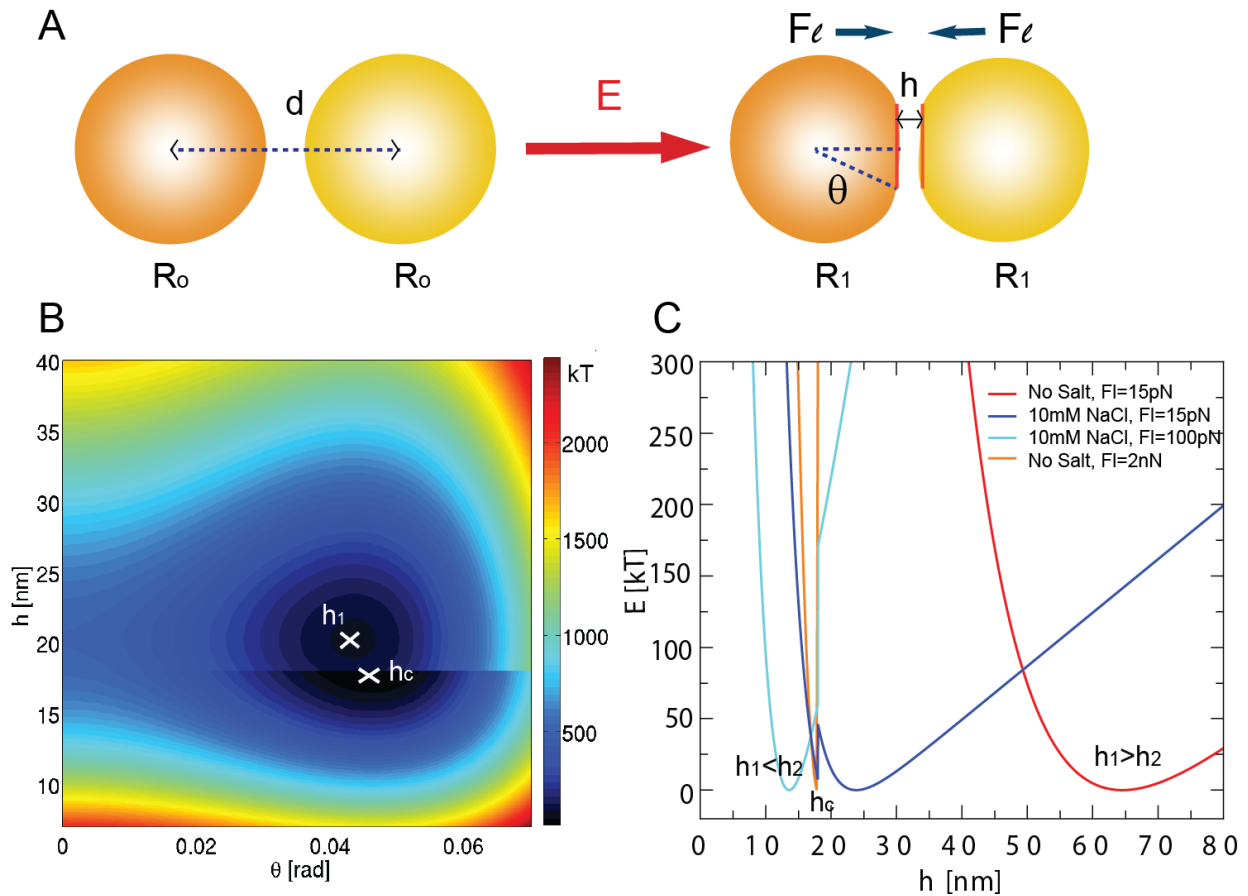

**Figure 2** (A) Idealized droplet deformation under a compressive force $F_l$. The total energy $E$ depends on the inter-distance $h$ and the deformation angle $\theta$. (B) A typical energy landscape as a function of the deformation parameters. Patch adhesion occurs when the global energy minimum is located at $h \leq hc$. (C) The model predicts that two droplets adhere spontaneously either by the addition of salt or an applied force.

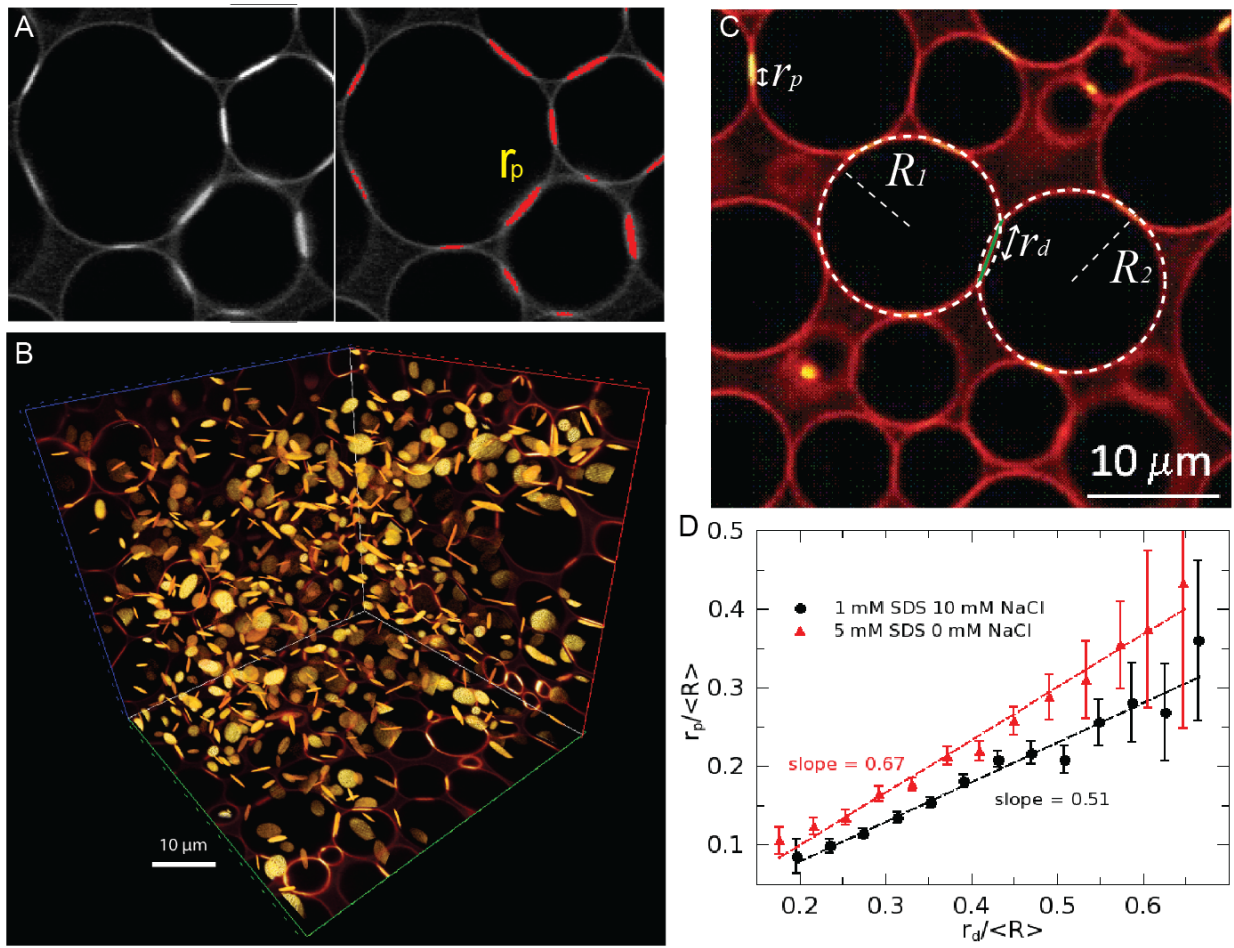

**Figure 3** (A) The radii of fluorescent adhesive patches are determined by thresholding the intensity of the images. (B) All patches are thus identified in the 3D structure of the packing. (C) The radii of deformation between droplets are derived from the overlap between identified spheres of radii R1 and R2. (D) This analysis reveals the linear correlation between the adhesion and deformation radii of each contacting droplet pair in the packing.

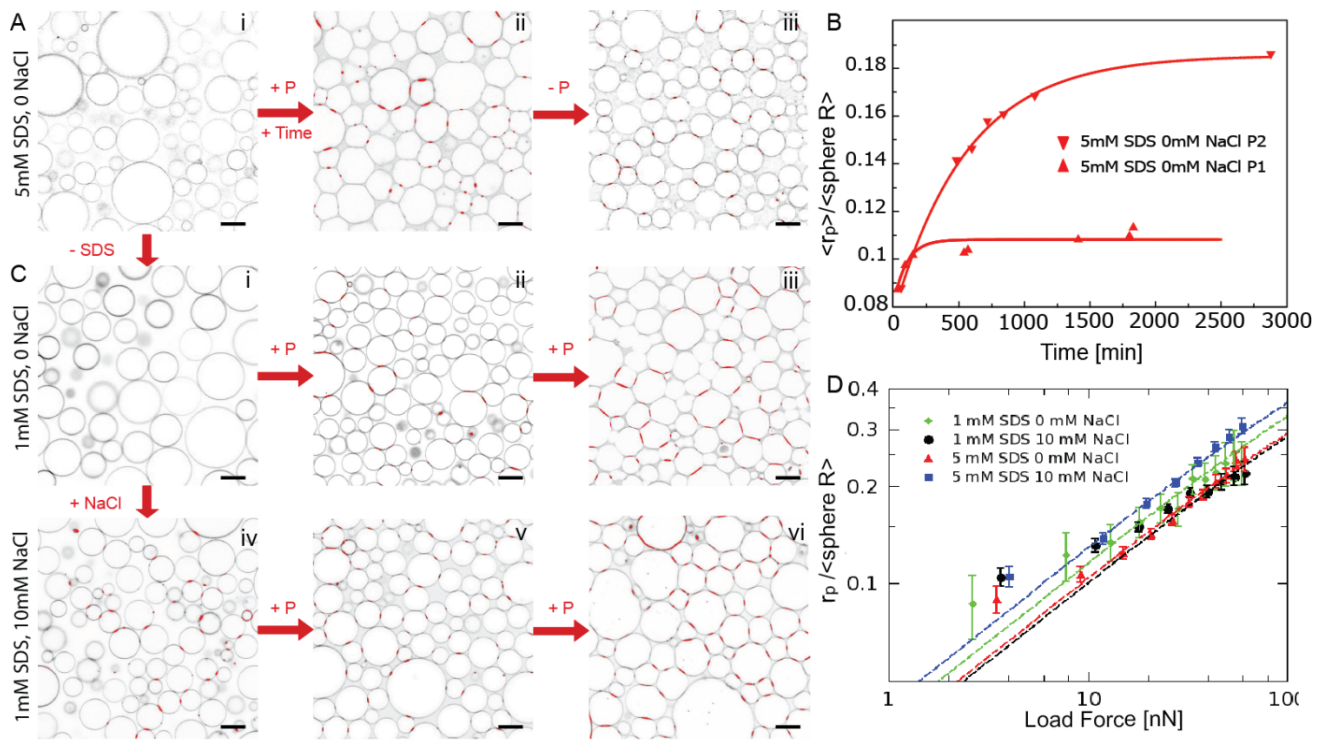

**Figure 4** Two-dimensional confocal slices are shown for different conditions. When the droplets are stabilized with 5mM SDS, centrifugation and waiting time are necessary to observe the formation of adhesive patches, which are irreversible (Ai-iii). (B) The growth of the mean adhesive radius for the emulsion in (A) is shown for two different applied compressions corresponding to average forces of <F1>= 26nN and <F2>= 42nN, and fit with exponentials (red lines). Lowering the SDS concentration to 1mM still requires compression to induce adhesion (Ci-iii). However, when salt is added to the solution (Civ-vi) gravity alone triggers droplet adhesion and the patches formed under compression are more numerous (Cvi) than in the no salt cases (Aiii, Ciii). (D) Normalized patch radius for all emulsion conditions grows as a function of the applied force, in good agreement with the model at high forces (dashed lines). Bars=10μm

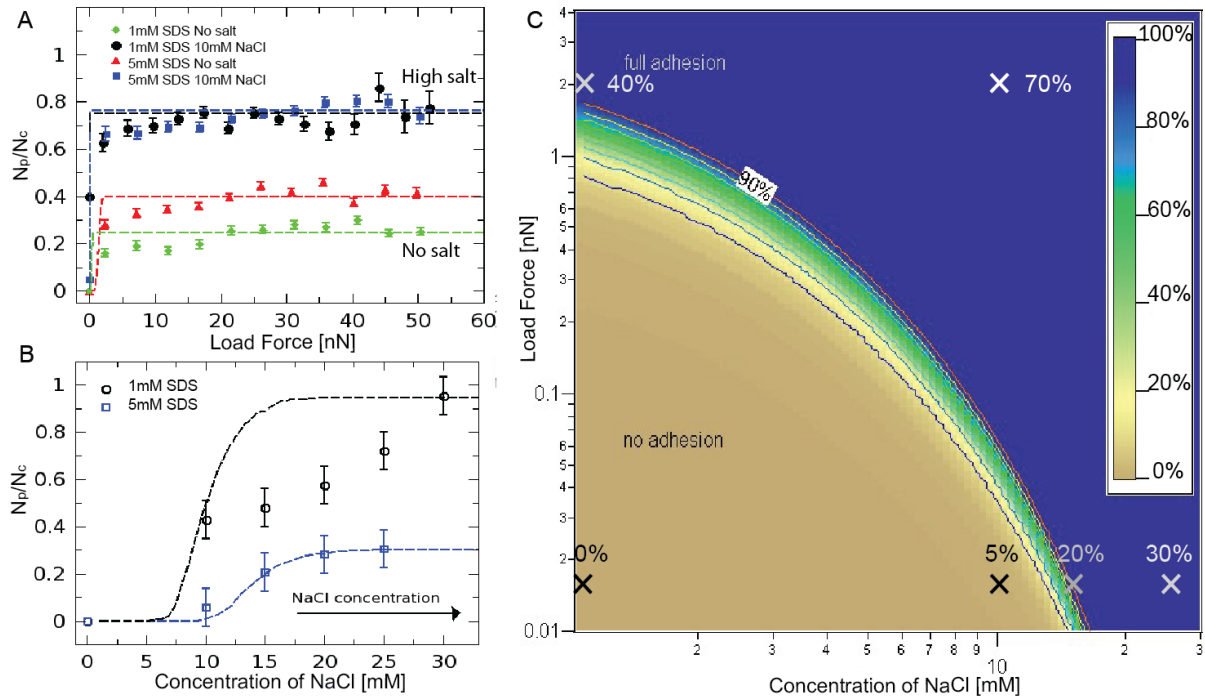

**Figure 5** The probability of finding a patch increases as a function of the applied force in (A) or salt concentration in (B), in agreement with the model (dashed lines). The observed trends are predicted by the model phase diagram of adhesion in (C), in which the parameters are consistent with the literature (see Supplementary Materials). The experimental fraction of adhesive contacts are labeled as crosses and overlaid with the model phase diagram.